\documentclass[10pt,aps,prl,preprintnumbers,superscriptaddress,showpacs,
twocolumn,floatfix,nofootinbib]{revtex4-1}
\usepackage{graphicx}
\usepackage{amsmath}
\usepackage{slashed}

\newcommand{\bfm}[1]{\mbox{\boldmath$#1$}}

\newcommand{\gsim}{\;\rlap{\lower 3.5 pt \hbox{$\mathchar \sim$}} \raise 1pt
\hbox {$>$}\;}
\newcommand{\lsim}{\;\rlap{\lower 3.5 pt \hbox{$\mathchar \sim$}} \raise 1pt
\hbox {$<$}\;}

\begin{document}

\title{\boldmath
Nonfactorizable QCD  Effects in Higgs Boson Production via Vector Boson  Fusion
\unboldmath}
\author{Tao Liu}
\email[]{ltao@ualberta.ca}
\affiliation{Department of Physics, University of Alberta, Edmonton, Alberta T6G
2J1, Canada}
\author{Kirill Melnikov}
\email[]{kirill.melnikov@kit.edu}
\affiliation{Institut f\"ur Theoretische Teilchenphysik,
 Karlsruher Institut f\"ur Technologie (KIT), 76128 Karlsruhe, Germany}
\author{Alexander A. Penin}
\email[]{penin@phys.ethz.ch}
\affiliation{Department of Physics, University of Alberta, Edmonton, Alberta T6G
2J1, Canada}
\affiliation{Institut f\"ur Theoretische Teilchenphysik,
Karlsruher Institut f\"ur Technologie (KIT), 76128 Karlsruhe, Germany}
\affiliation{Institute for Theoretical Physics, ETH Z\"urich, 8093 Z\"urich,
Switzerland}
\begin{abstract}
We discuss  nonfactorizable  QCD corrections to  Higgs boson production in
 vector boson fusion at the Large Hadron Collider. We point out
that these corrections can be  computed in the eikonal approximation
retaining all the terms that are not suppressed by  the ratio of the
transverse momenta of the tagging jets  to the  total center-of-mass energy.
Our analysis shows  that in certain kinematic distributions the
nonfactorizable corrections can be as large as  a percent  making them
quite comparable to their factorizable  counter-parts.
\end{abstract}
%\pacs{11.15.Bt, 12.38.Bx, 12.38.Cy}
\preprint{ALBERTA-THY-04-19, TTP19-017, P3H-19-013}

\maketitle
Vector boson fusion (VBF) is one of the two key channels for  Higgs boson
production at the Large Hadron Collider (LHC)
\cite{Khachatryan:2015bnx,Aaboud:2018gay}. Studies of  Higgs boson
properties  in this process require  accurate theoretical
prediction for its cross section and kinematic distributions.  Radiative corrections, both
QCD and electroweak, are important for the reliable description of these processes.
Current understanding of  QCD corrections to the VBF Higgs boson production is
highly  advanced: following  the original calculation  of the next-to-leading
(NLO)  corrections \cite{Figy:2003nv}, both the next-to-next-to-leading
(NNLO) \cite{Bolzoni:2010xr,Cacciari:2015jma,Cruz-Martinez:2018rod} and the
next-to-next-to-next-to-leading (N$^3$LO)  \cite{Dreyer:2016oyx} corrections
were computed in the so-called structure function approximation \cite{Han:1992hr}.
The electroweak corrections to VBF were computed in Ref.~\cite{cicco}. Other
interesting effects such as loop-induced interference between Higgs production
in gluon fusion and in vector boson fusion,   and  the gluon-initiated VBF Higgs production were
studied in Refs.~\cite{jeppe,rob}, respectively.

The structure function approximation --
the centerpiece  of the current studies  of QCD effects in VBF --
neglects interactions between incoming
QCD partons and retains  QCD effects confined to a single fermion line. There
are good reasons for doing this. Indeed,  at NLO the gluon exchanges between
different quark lines do not change the VBF cross section  as a consequence
of  color conservation. At NNLO, two gluons  exchanged between two
fermion lines can be in a color-singlet state and for this reason do
contribute to the VBF cross section.  Such nonfactorizable corrections,
however, are necessarily color-suppressed,  making it plausible that they are
small. This argument was used as the  justification  for computing higher-order QCD
corrections to  VBF  Higgs boson production in the
structure function approximation \cite{Bolzoni:2010xr}.

However, it is  interesting to ask just for  how long does it  make sense to
improve the precision on the factorizable contributions while ignoring  the
nonfactorizable ones. This question appears to be quite relevant  since
computations of factorizable contributions have advanced to very high orders
in perturbative QCD \cite{Dreyer:2016oyx}.  Answering this question is
difficult  since  not much is known about nonfactorizable corrections beyond
their color suppression. As we already mentioned, these corrections  do not
contribute at NLO while at NNLO they require  two-loop five-point
functions that depend on many kinematic variables and  the masses of vector
bosons and the Higgs boson.  Thus, the technical complexity of perturbative
computations  required to obtain the two-loop nonfactorizable contribution
appears to be  overwhelming to expect significant advances in the foreseeable
future.  An estimate of nonfactorizable corrections that makes use of QCD
dynamics and in this sense  goes beyond the color-suppression argument is
highly desirable, in our opinion.

In this Letter we will show that it is possible and in fact rather  simple to
compute the dominant contribution to  nonfactorizable corrections,  making
use of the particular kinematics of the VBF process. Indeed, this process is
identified by the presence of two forward tagging jets whose transverse
momenta are  small compared to their energies. Thus, we can try to compute
the nonfactorizable corrections in an approximation where we only retain
contributions that are leading in $p_{j,\perp}/\sqrt{s}$, where $p_{j,\perp}$
is a  transverse  momentum of a tagging jet,  and $s$ is the
center-of-mass energy squared of the colliding partons. Since in the VBF
process  $\sqrt{s} \gsim 600~{\rm GeV}$ and $p_{j,\perp} \sim 100~{\rm GeV}$
\cite{Khachatryan:2015bnx,Aaboud:2018gay} this approximation is justified in
 large part of the phase space.

\begin{figure}[t]
\begin{center}
\begin{tabular}{ccc}
\hspace*{-5mm}\includegraphics[width=3.6cm]{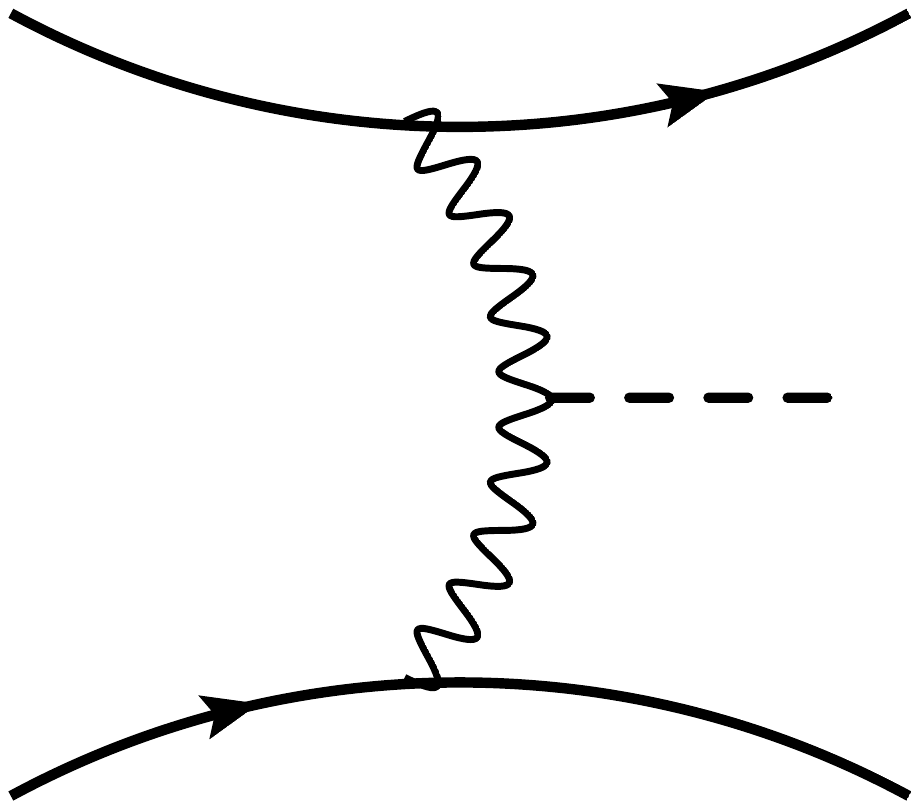}&
\hspace*{-8mm}\includegraphics[width=3.6cm]{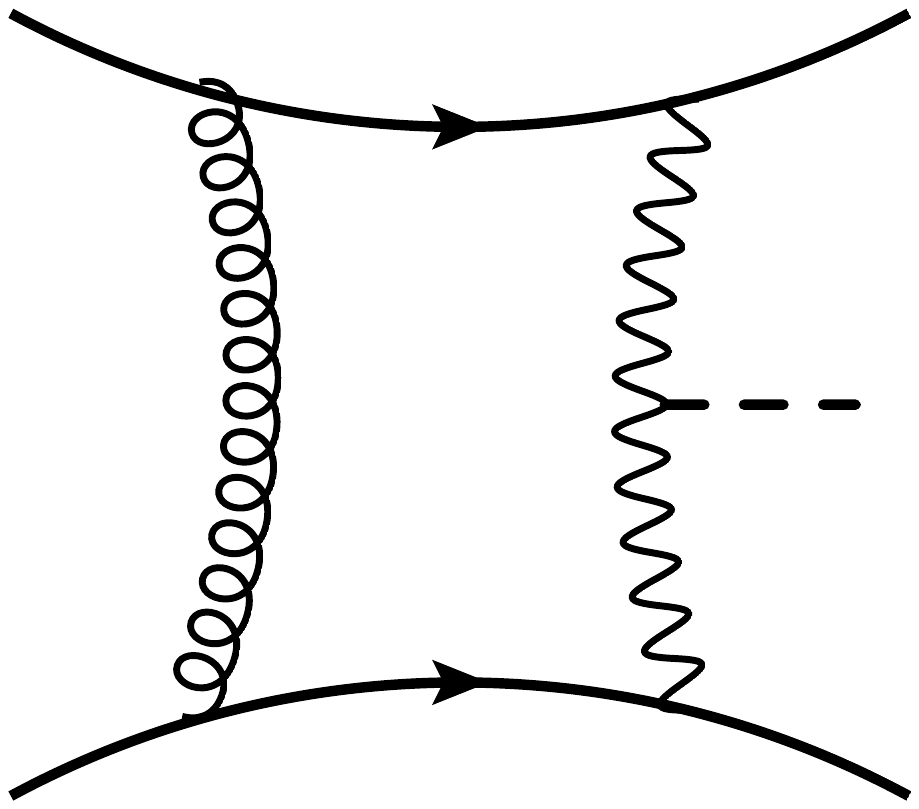}&
\hspace*{-8mm}\includegraphics[width=3.6cm]{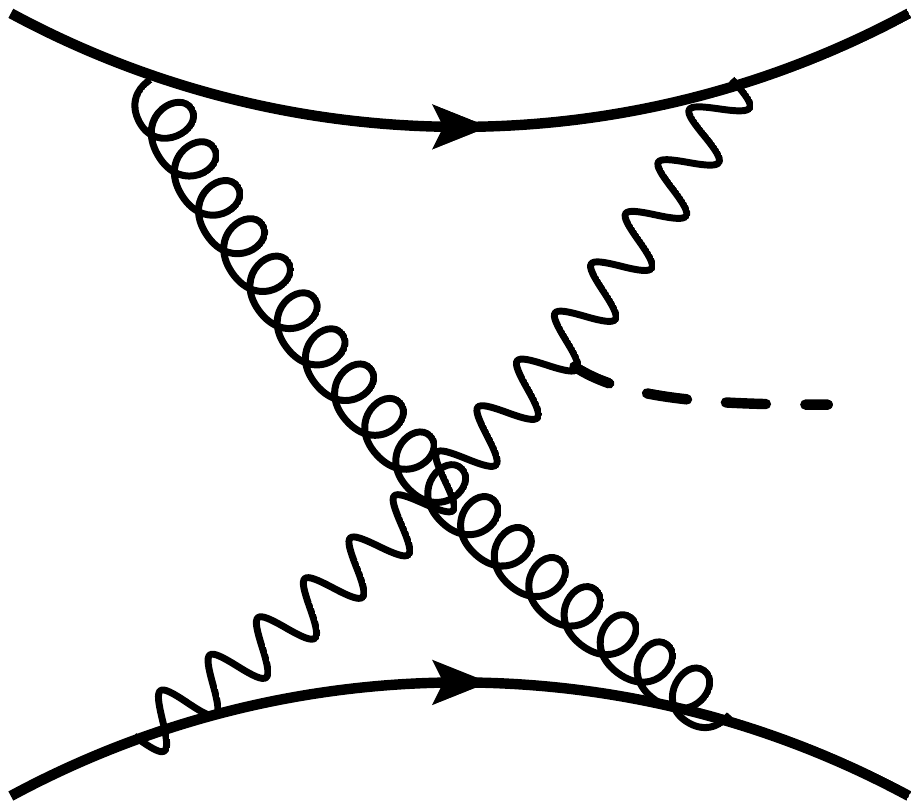}\\[-22mm]
\hspace*{-3mm}(a)&\hspace*{-5mm}(b)&\hspace*{-5mm}(c)\\
\end{tabular}
\end{center}
\caption{\label{fig::1} The Feynman diagrams for the Higgs boson production
in VBF: (a) the Born amplitude, (b,\,c) the one-loop
nonfactorizable QCD corrections. The solid, dashed, wavy and loopy lines
stay for quark, Higgs, vector boson and gluon fields, respectively.}
\end{figure}

It is well known that  the computation of cross sections   at  leading
power in the small ratio $p_{\perp,j}/\sqrt{s}$,  can be performed within the
eikonal approximation for the colliding  particles
\cite{Cheng:1970jk,Chang:1969by,Lipatov:1976zz}.  To explain this
approximation, we consider a collision of two quarks that leads to the
production of the Higgs boson in VBF
\begin{equation}
q_1(p_1)+ q_2(p_2) \to q_1(p_3) + q_2(p_4) + H(p_5).
\nonumber
\end{equation}
The  leading order contribution to this  process is  shown in
Fig.~\ref{fig::1}(a). The eikonal approximation  separates dynamics in the
plane spanned by the two four-momenta of the incoming quarks $p_{1,2}$ from
dynamics  in the plane that is transversal to it. We will refer to a
component of a four-vector $k^\mu$ in the transversal plane as $k_\perp^\mu$
or ${\bfm k}$. We choose the reference frame in such a way that $p_1$ and
$p_2$ have only a single light-cone component $p^-_1$ and $p^+_2$,
respectively. Then in the eikonal approximation a gauge boson coupling to the
quark line with momentum  $p_1$ ($p_2$) is obtained by replacing the
corresponding current $j^\mu$ with its light-cone component $j^-$ ($j^+$)
while the quark propagators are replaced as follows
\begin{equation}
  \begin{split}
  &{1\over \slashed{p}_{1,2}+\slashed{k}+i\epsilon}
    \to {\gamma^{\pm}\over 2k^{\pm}+i\epsilon}
    \,,
%\\
%  &{1\over \slashed{p}_2+\slashed{k}+i\epsilon}
%  \to {\gamma^-\over 2k^-+i\epsilon}\,.
  \end{split}
  \label{eq::prop}
\end{equation}
where $\gamma^{\pm}$ are the light-cone components of the Dirac $\gamma$-matrices.

In the VBF process Higgs  bosons are  produced at central rapidities so that
they are well-separated from the tagging jets. This ensures that  momentum
transfers $q_3=p_3-p_1$ and $q_4=p_4-p_2$ mostly  have transverse components
${\bfm q_{3,4}}$ while their light-cone component are  suppressed by $p_{{3,4},\perp}/\sqrt{s}$.
Thus,  the Higgs boson emission does not spoil the applicability of
the eikonal approximation.

\begin{figure}[t]
\begin{center}
\begin{tabular}{cc}
\includegraphics[width=3.6cm]{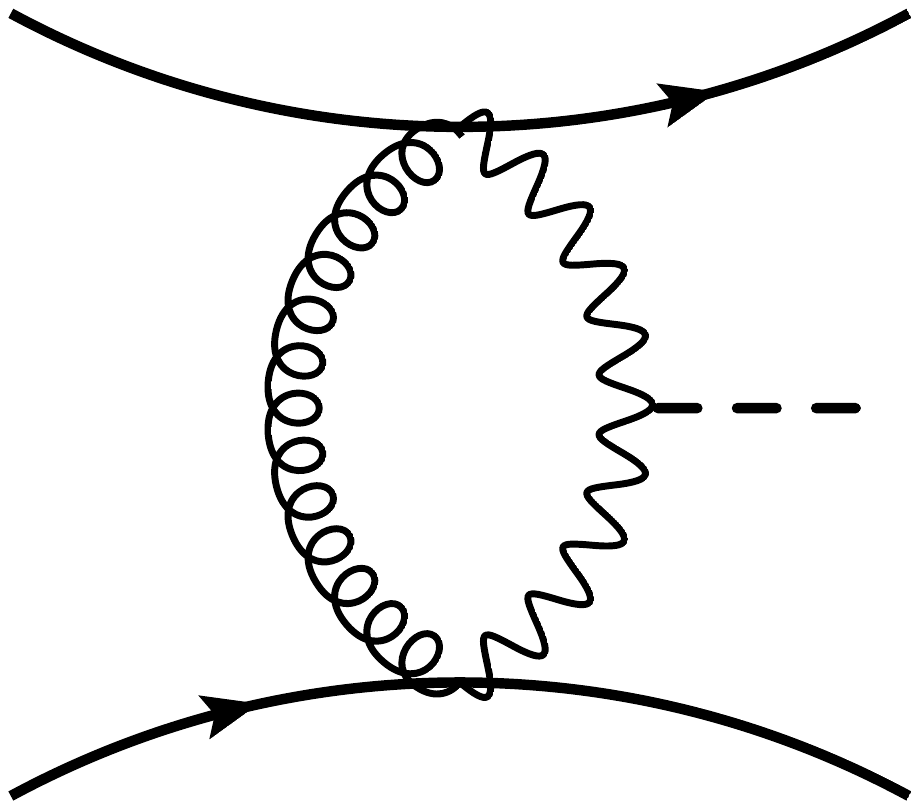}&
\hspace*{-5mm}\includegraphics[width=3.6cm]{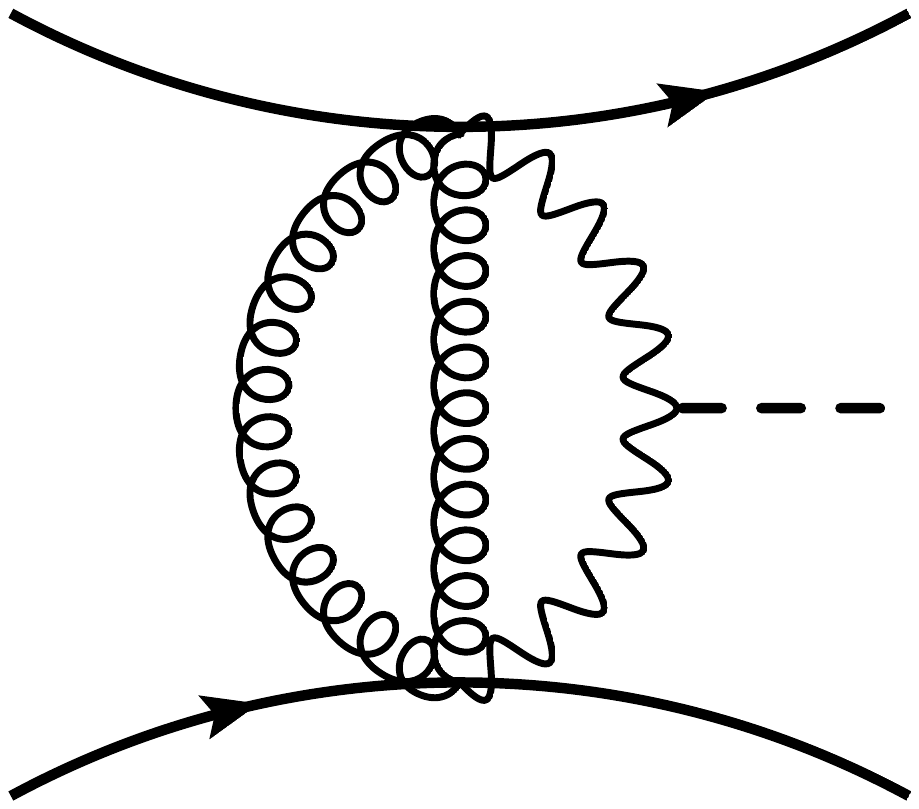}\\[-22mm]
\hspace*{2mm}(a)&\hspace*{-3mm}(b)\\
\end{tabular}
\end{center}
\caption{\label{fig::2} One- and two-loop transversal space Feynman diagrams.}
\end{figure}

We continue with the discussion of the nonfactorizable QCD corrections. In
the one-loop approximation the relevant diagrams are shown  in
Fig.~\ref{fig::1}(b,c). Since electroweak vector bosons do not carry color,
the one-loop contribution to the cross section vanishes at NLO by  color
conservation. Nevertheless,  the  square of the one-loop amplitude  contributes to the
NNLO  cross section along with the generic two-loop nonfactorizable
corrections. In both cases the two gluons connecting the different quark
lines must be in a  color-singlet configuration. Thus we can compute the
corrections  by replacing  gluons by  abelian gauge bosons with the effective
coupling $\tilde{\alpha}_s=\left({N_c^2-1\over
4N_c^2}\right)^{1/2}\!\!\!\alpha_s$, where $N_c=3$ and the prefactor arises
from averaging over  colors. Considering the sum of the planar and non-planar
diagrams in Figs.~\ref{fig::1}(b,c), we find that the eikonal quark
propagators  add up to $1/(2 k^\pm+i\epsilon)-c.c. =-i\pi\delta(k^\pm)$.
Hence, when the two diagrams are combined, the virtual quark propagators are
replaced by $\delta(k^+)$ and $\delta(k^-)$ and the light-cone dynamics
decouples. Thus the computation of the nonfactorizable one-loop contribution
is reduced to the analysis  of the effective Feynman diagram shown in
Fig.~\ref{fig::2}(a)  in the {\it two-dimensional transversal space}. In  the
eikonal approximation  QCD corrections are diagonal in the chiral  basis.
This implies that (for a given type of electroweak  gauge bosons that fuse
into the Higgs) the Born amplitude ${\cal M}^{(0)}$  factors out. Hence, the
expression for the one-loop amplitude  can be written as follows
\begin{equation}
  \begin{split}
   & {\cal M}^{(1)}  =
   i \tilde{\alpha}_s
   \chi^{(1)}( {\bfm q}_3, {\bfm q}_4){\cal M}^{(0)}\,,
   \end{split}
   \label{eq::M1}
\end{equation}
with
\begin{equation}
  \begin{split}
  & \chi^{(1)}( {\bfm q}_3, {\bfm q}_4) =  \frac{1}{\pi}
  \int \frac{{\rm d}^2  {\bfm k} }{   {\bfm k}^2 + \lambda^2} \\
  & \times \frac{{\bfm q}_3^2 +
  M_V^2}{ ({\bfm k} - {\bfm q}_3)^2 + M_V^2}
  \frac{{\bfm q}_4^2 + M_V^2}{ ({\bfm k} + {\bfm q}_4)^2 + M_V^2  }\,,
  \end{split}
  \label{eq::chi1}
\end{equation}
where   $M_V = M_{Z,W}$ is an electroweak boson mass. We note that the
function $\chi^{(1)}$ is  ultraviolet-finite but  infrared-divergent. To
regulate the infrared divergence, we introduced an auxiliary gluon mass
$\lambda$. Moreover, the function $\chi^{(1)}$ is explicitly real, so that
the entire one-loop correction is  imaginary. This is  yet another reason,
in addition to color conservation,  that leads to vanishing  interference
between the one-loop amplitude computed in the eikonal approximation and the
leading order amplitude.

At  two loops the structure of the corrections is similar. In the
color-singlet configuration the gluon vertices commute and the factorization
property of the eikonal approximation \cite{Sudakov:1954sw} can be applied. As
a result  the sum over all the permutations of the gluon and  vector-boson
vertices reduces to the effective transversal space diagram in
Fig.~\ref{fig::2}(b) \cite{Cheng:1970jk}. The corresponding expression for
the amplitude reads
\begin{equation}
  \begin{split}
  & {\cal M}^{(2)} =- {\tilde{\alpha}^2_s\over 2!}
  \chi^{(2)}( {\bfm q}_3, {\bfm q}_4){\cal M}^{(0)}\,,
  \end{split}
  \label{eq::M2}
\end{equation}
where $1/2!$ factor results from the symmetrization of two   identical gluons and
\begin{equation}
  \begin{split}
  & \chi^{(2)}( {\bfm q}_3, {\bfm q}_4) =\frac{1}{\pi^2}
  \int  \left(\prod \limits_{i=1}^{2}
  \frac{ {\rm d}^2  {\bfm k}_i}{   {\bfm k}_i^2 + \lambda^2 }\right) \\
  & \times \frac{{\bfm q}_3^2 +
  M_V^2}{ ({\bfm k}_1+{\bfm k}_2 - {\bfm q}_3)^2 + M_V^2}
  \frac{{\bfm q}_4^2 + M_V^2}{ ({\bfm k}_1+{\bfm k}_2
  + {\bfm q}_4)^2 + M_V^2  }\,.
  \end{split}
  \label{eq::chi2}
\end{equation}
Squaring the sum of tree-, one- and two-loop contributions to the scattering
amplitude, we obtain the NNLO QCD correction to the cross section due to
nonfactorizable contributions
\begin{eqnarray}
 {\rm d}\sigma^{\rm NNLO}_{\rm nf}&=&
 \left({N_c^2-1\over 4N_c^2}\right)\alpha_s^2\,
 \chi_{\rm nf}\,
 {\rm d}\sigma^{\rm LO}\,.
 \label{eq::sigma}
\end{eqnarray}
In Eq.(\ref{eq::sigma}) ${\rm d} \sigma^{\rm LO}$ is the leading-order differential cross
section for VBF and
\begin{equation}
 \chi_{\rm nf}({\bfm q_3},{\bfm q_4})
 =\left[\chi^{(1)}({\bfm q_3},{\bfm q_4})\right]^2 -
 \chi^{(2)}({\bfm q_3},{\bfm q_4})\,
\label{eq::chinf}
\end{equation}
is the nonfactorizable correction.

The nonfactorizable correction has peculiar properties.
It is
independent of  the vector boson couplings to quarks and to the Higgs boson;
these couplings are accommodated  in the leading order cross section in
Eq.~(\ref{eq::sigma}). In the large-$N_c$ limit the color factor in
Eq.~(\ref{eq::sigma}) remains finite while for the factorizable corrections
it grows as $N_c^2$ providing the color suppression of the nonfactorizable
contribution. Finally, the two terms in Eq.~(\ref{eq::chinf}) are separately infrared
divergent. These divergences, however, are not related to the usual (nonfactorizable) real soft
gluon emissions that, in fact, are suppressed as $p_\perp/\sqrt{s}$ and, therefore,  do not
contribute  to the VBF cross section at leading power.
The infra-red divergencies in non-factorizable corrections  originate from the exchange of static {\em
Glauber} gluons \cite{Glauber} propagating in the transversal space.  It is
well known that  when  abelian gauge bosons are  exchanged,
the amplitudes acquire a factor $e^{i\phi}$ where $\phi$ is  the
infrared-divergent Glauber phase $\phi=-\tilde{\alpha}_s\ln\lambda^2$
\cite{Cheng:1970jk}. This phase factor disappears  in the cross section,
which means that the infrared-divergent parts of the first and the second
term in Eq.~(\ref{eq::chinf}) must cancel each other.

To show this cancellation explicitly,  we  consider
the $\lambda \to 0$ limit, extract the infrared singularities
from the two  functions $\chi^{(1,2)}$  and write them as follows
  \begin{equation}
    \begin{split}
    &  \chi^{(1)}
    = -\ln \left(\frac{\lambda^2}{M_V^2}\right) + f^{(1)}\,, \\
    &  \chi^{(2)} = \ln^2\left( \frac{\lambda^2}{M_V^2}\right)
    - 2 \ln \left(\frac{\lambda^2}{M_V^2}\right) f^{(1)} + f^{(2)}\,.
  \end{split}
  \label{eq::chirep}
  \end{equation}
The functions $f^{(1),(2)}$ read
\begin{eqnarray}
  f^{(1)} &= &\int \limits_{0}^{1} {\rm d} x \frac{\Delta_3 \Delta_4}{r^2_{12}}
  \left [ \ln \left(\frac{r_{12}^2}{r_2 M_V^2}\right) + \frac{r_1-r_2}{r_2} \right ],
  \nonumber\\
   f^{(2)} &=& \int \limits_{0}^{1} {\rm d} x \frac{\Delta_3 \Delta_4}{r_{12}^2}
  \left[\left(\ln \left(\frac{r_{12}^2}{r_2 M_V^2}\right) + \frac{r_1-r_2}{r_2} \right)^2
     \right.
     \label{eq::frep} \\
  & & \left.
  - \ln^2 \left ( \frac{r_{12}}{r_2} \right )
  -\frac{2r_{12}}{r_2}\ln \left ( \frac{r_{12}}{r_2} \right )
   -2\,{\rm Li}_2\left({r_1\over r_{12}}\right)
    \nonumber \right.  \\
    && \left.
    -\left(\frac{r_1-r_{2}}{r_2}\right)^2+\frac{\pi^2}{3}
    \right],
  \nonumber
\end{eqnarray}
where  we used  the notations
  \begin{equation}
\begin{split}
   &  r_1 = {\bfm q}_3^2 \,x + {\bfm q}_4^2\,(1-x)
   - {\bfm q}_H^2 \,x (1-x)\,,\\
   &  r_2 = {\bfm q}_H^2\, x(1-x) + M_V^2\,,\\
   &  r_{12} = r_1 + r_2\,,\\
   & \Delta_i = {\bfm q}_i^2 + M_V^2\,.
  \end{split}
  \label{eq::rdef}
  \end{equation}
In Eq.(\ref{eq::rdef}) ${\bfm q}_H=-{\bfm q}_4-{\bfm q}_3$ is the Higgs boson transverse
momentum.  This result can be obtained by using the Feynman parameter
representation for one- and two-loop two-dimensional triangle diagrams
corresponding to the functions $\chi^{(1)}$ and $\chi^{(2)}$ respectively.

We note that it should be possible to compute the two functions
analytically.\footnote{It is well-known that in the two-dimensional
space-time three-point functions can be described by linear combinations of
two-point functions. The one-loop case is explicitly discussed in
Ref.~\cite{Ellis:2011cr}.} However, the one-dimensional integral
representations in Eqs.~(\ref{eq::frep},\ref{eq::rdef}) are perfectly
suitable for the numerical evaluation of the nonfactorizable corrections so
that we decided not to pursue the analytic calculation further.

Using representations  Eq.~(\ref{eq::chirep}) in Eq.~(\ref{eq::chinf}), we obtain
the  finite result
\begin{equation}
\chi_{\rm nf}({\bfm q_3},{\bfm q_4})  = [f^{(1)}({\bfm q_3},{\bfm q_4})]^2-f^{(2)}({\bfm q_3},{\bfm q_4})\,,
\label{eq::chifin}
\end{equation}
for the two-loop  nonfactorizable correction to the VBF cross section. It can
be  used for the numerical evaluation of the correction factor for values of
the transverse momenta that are much smaller than the energy of the two
colliding partons.

It is instructive  to compute the function $\chi_{\rm nf}$ in a few limiting
cases.  The simplest case is when all the transverse momenta are small
compared to the vector boson mass $|{\bfm q}_{3,4}|\ll M_V$. In this limit
$r_1 = 0$, $r_2 = M_V^2$ and we find
\begin{equation}
  \lim_{q_{3,4} \to 0}   \chi_{\rm nf}  = 1- \frac{\pi^2}{3}\,.
  \label{eq::lim0}
  \end{equation}
Another  interesting case is when the Higgs boson momentum is small  ${\bfm
q}_H^2\ll M_V^2,{\bfm q}_{3,4}^2$. In this limit $r_1 = {\bfm q}^2_{3}$, $r_2
= M_V^2$ and  we obtain
\begin{equation}
  \begin{split}
 & \lim_{q_H \to 0} \chi_{\rm nf}  =
 \ln^2\left({1+x\over x}\right)+2\,{\rm Li}_2\left({1\over 1 + x}\right)-
 \frac{\pi^2}{3}\\
 & + 2\,{1+x\over x}\ln\left({1+x\over  x}\right)+\left({1-x\over x}\right)^2,
   \end{split}
 \label{eq::lim1}
\end{equation}
with  $x=M_V^2/{\bfm q}_3^2$. In the opposite limit  when the transverse
momentum ${\bfm q}_3$ of one of the  tagging jets  is small compared to ${\bfm
q}_4\approx {\bfm q}_H$  the result reads
\begin{equation}
 \lim_{q_3 \to 0} \chi_{\rm nf}  =\ln^2\left({1+x\over  x}\right)
 +2\,{\rm Li}_2\left({1\over 1 + x}\right)-
 \frac{\pi^2}{3}\,.
\label{eq::lim2}
\end{equation}
The coefficient of the quadratic  logarithm in Eq.~(\ref{eq::lim1}) can
be read off from the infrared divergences of the one- and two-loop massless
amplitudes at zero Higgs boson momentum.   We have verified this coefficient
by exact evaluation of the scattering amplitudes in dimensional regularization as
functions of ${\bfm q}_3^2/s$  with subsequent expansion of the result at
small transverse momentum; this calculation provides   a nontrivial test of the eikonal
approximation used in the above analysis.

%----------------------------------------------------------------------
\begin{figure*}
  \centering
          \includegraphics[clip,width=5.5cm,page=1,angle=0]{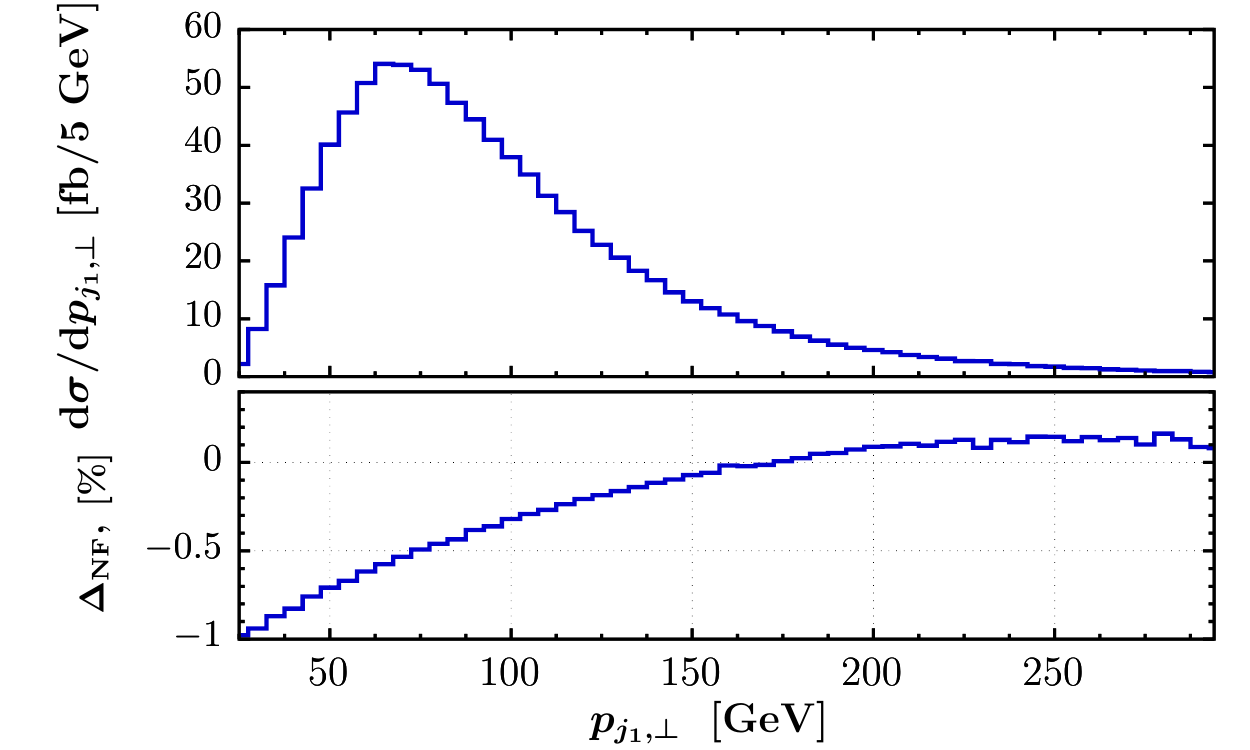}%
          \hfill\includegraphics[clip,width=5.5cm,page=1,angle=0]{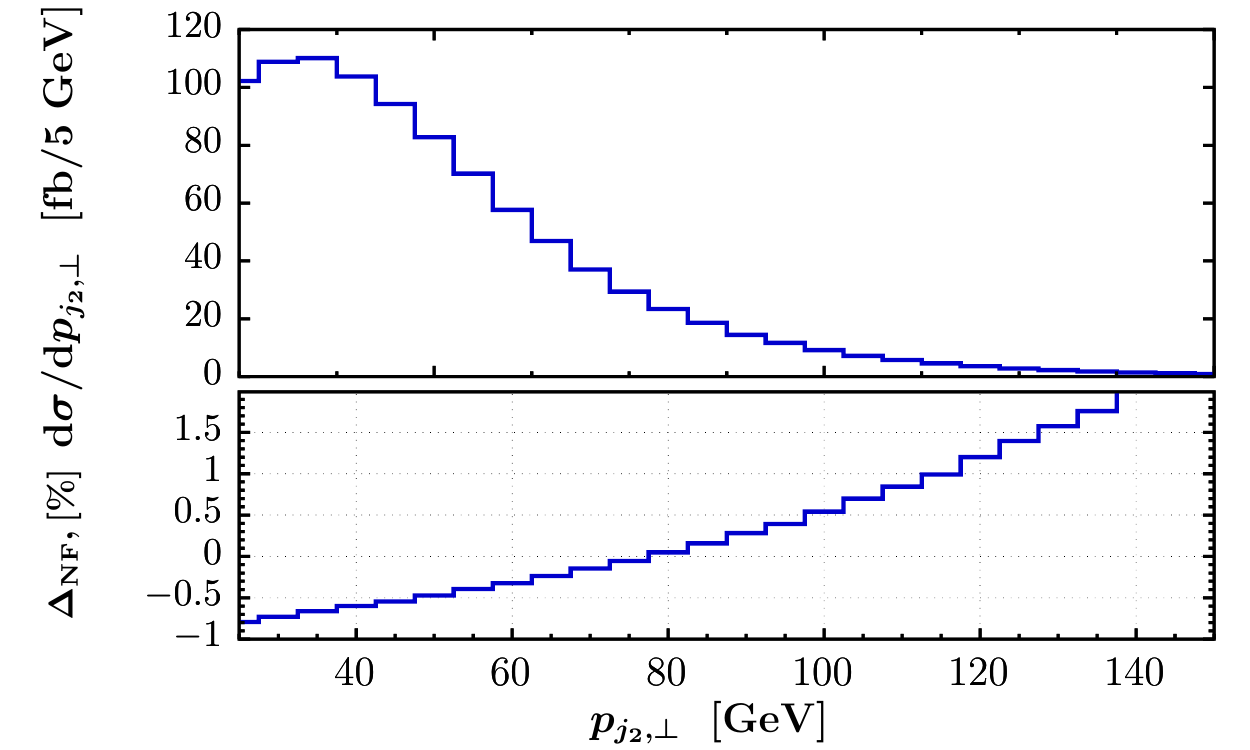}%
      \hfill\includegraphics[clip,width=5.5cm,page=1,angle=0]{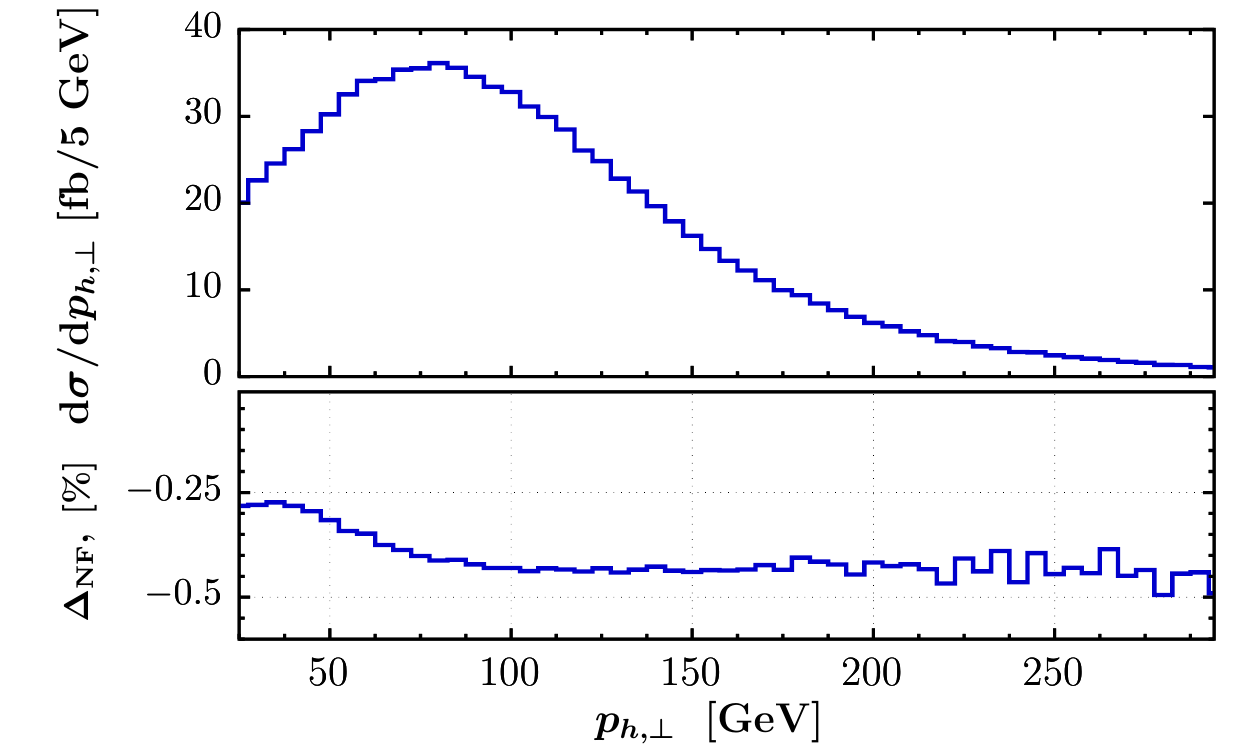}\hspace{0.8mm}%
              \includegraphics[clip,width=5.5cm,page=1,angle=0]{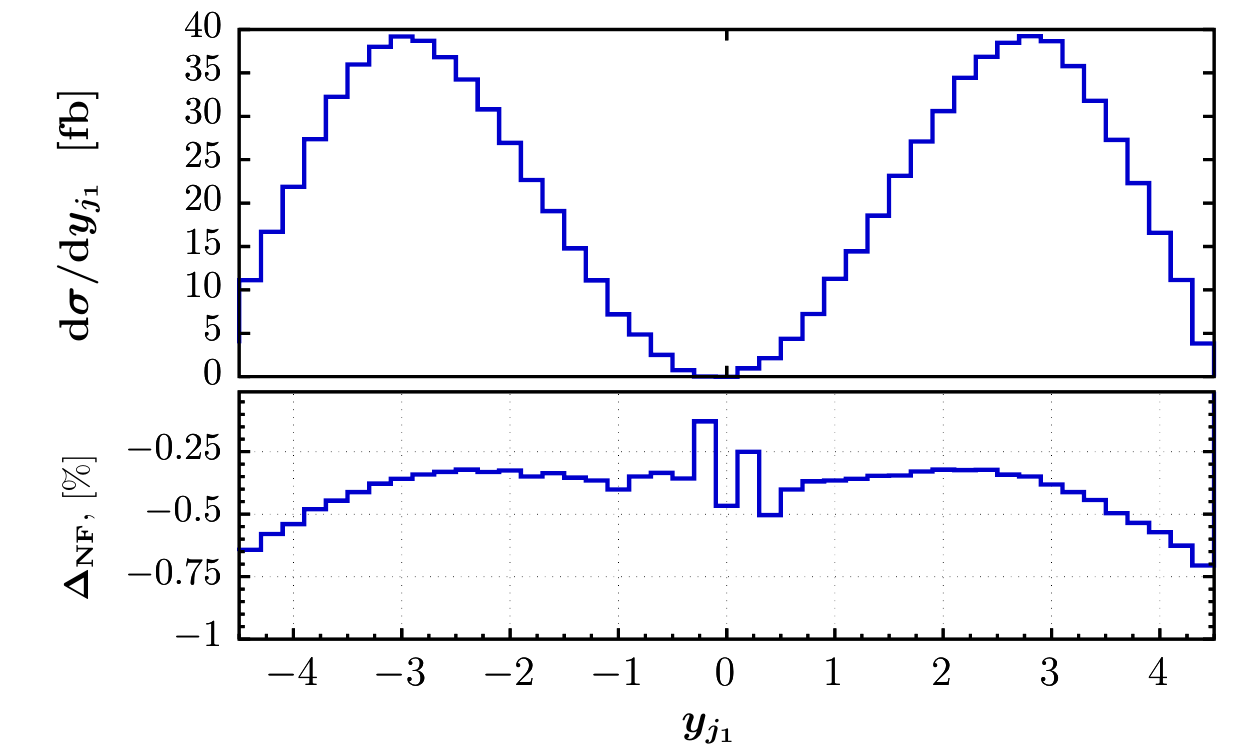}%
    \hfill\includegraphics[clip,width=5.5cm,page=1,angle=0]{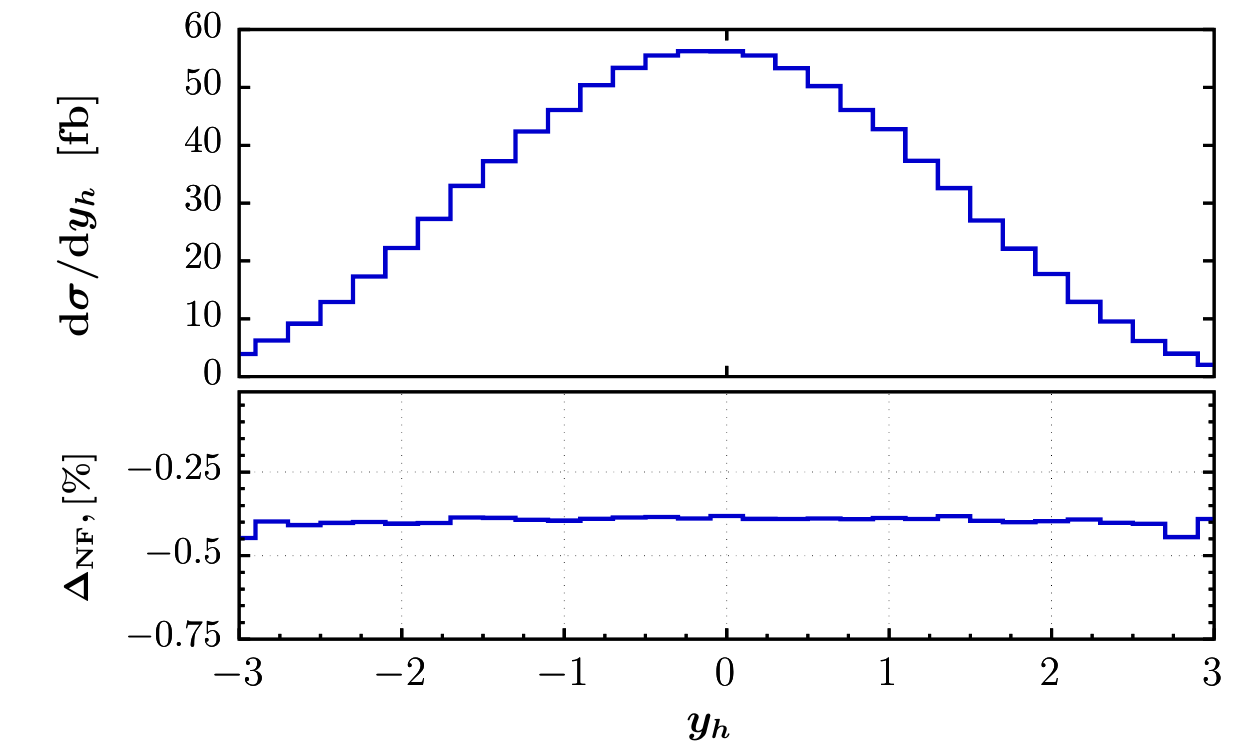}%
    \hfill\includegraphics[clip,width=5.5cm,page=1,angle=0]{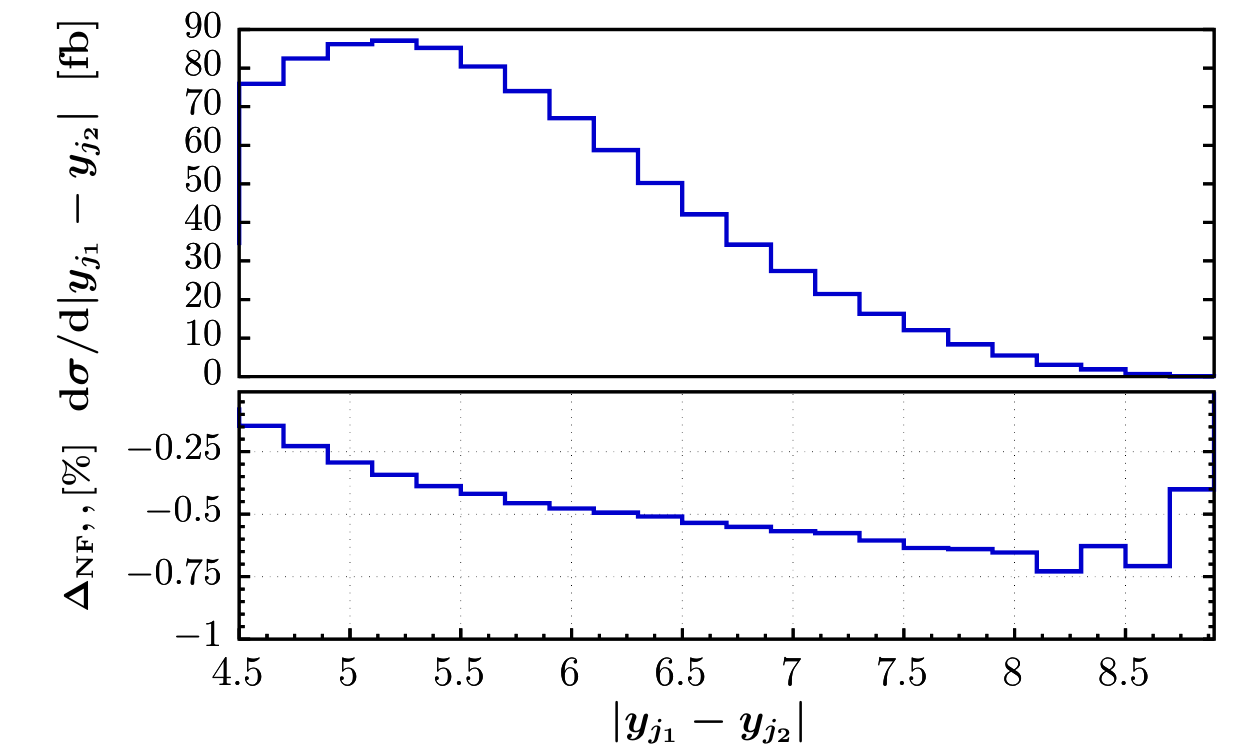}\hspace{0.8mm}%

\caption{Differential cross sections for the transverse momentum
distributions  for the two leading jets, $p_{j_1,\perp}$ and $p_{j_2,\perp}$
and  for the Higgs boson, $p_{h,\perp}$ as well as the rapidity distributions
of the first jet and of the Higgs boson in VBF at the $13~{\rm TeV}$ LHC.  We also show  the rapidity
difference distribution of the two tagging jets.  See text for details. }
  \label{fig::3}
\end{figure*}
%----------------------------------------------------------------------

One can use these asymptotic formulas to discuss characteristic features of
the nonfactorizable contribution. For example,
taking $\alpha_s \sim 0.1$,   we find that  the zero-momentum limit Eq.~(\ref{eq::lim0})
implies minus one  percent correction to the differential cross section
Eq.~(\ref{eq::sigma}).  At the same time, the limit of the small Higgs
transverse momentum Eq.~(\ref{eq::lim2}) suggests that {\it positive}
corrections as large as a few percent occur when the transverse momentum of
the tagging jets exceeds $100~{\rm GeV}$.

It follows from the above discussion that  the  nonfactorizable corrections
can reach a  few percent in differential distributions. However, since the
corrections appear with opposite  signs at low and high transverse momenta,
they may cancel in quantities that are inclusive with respect to  kinematic
features of the tagging jets.

The reason behind  sizable  nonfactorizable effects can be traced to their
connection to  the Glauber scattering phase. This connection leads to a
$\pi^2$-enhancement of the nonfactorizable contribution characteristic to the
imaginary phase, which partially overcomes the effect of the color
suppression, {\it cf.}  Eq.~(\ref{eq::sigma}). Interestingly,   previous attempts
to estimate nonfactorizable corrections were based on the
analysis of  real radiation \cite{Figy:2007kv} or the {\it real} part of
the one-loop amplitude \cite{Bolzoni:2011cu} which are insensitive to
contributions of this type.

Having discussed  features of the nonfactorizable contribution, we can now
evaluate its impact on the VBF Higgs production cross section.  We consider
proton-proton collisions at the LHC with the center of mass energy $13~{\rm TeV}$.
To select VBF
events, we require that tagging jets have transverse momenta larger than
$25~{\rm GeV}$ and   their invariant mass exceeds $ 600~{\rm GeV}$.  Besides
that, jets' rapidities should satisfy the conditions $y_{j_1,j_2} \in
[-4.5,4.5]$ and $|y_{j_1} - y_{j_2}| > 4.5$. In addition,  jets are
required to be in  opposite hemispheres. To compute the leading order
cross section and the nonfactorizable corrections we adopt the following
factorization and renormalization scales
\begin{equation}
\mu_F = \left [ \frac{m_h}{2} \sqrt{ \frac{m_h^2}{4}
+ p_{h,\perp}^2} \right ]^{1/2},\;\;\;
\mu_R = \sqrt{p_{j_1,\perp} p_{j_2,\perp}}\,.
\label{eq:scales}
\end{equation}
Note that our choice of the  factorization scale is identical to that of
Ref.~\cite{Cacciari:2015jma} which ensures that our leading order cross
sections and kinematic distributions are in agreement with that references.
For numerical simulations we use the NNPDF 3.0 parton distribution function
(\verb|NNPDF_nnlo_as_0118|) with the default value  $\alpha_s(m_Z) = 0.118$.
Electroweak parameters are determined from the Fermi constant $G_F = 1.16637
\times 10^{-5}~{\rm GeV}^{-2}$ and masses of electroweak gauge bosons $M_W =
80.398~{\rm GeV}$ and $M_Z = 91.1876~{\rm GeV}$. We take the mass of the
Higgs boson to be $m_H = 125~{\rm GeV}$. Within the above setup we obtain
the VBF cross section and the non-factorizable contribution at the  $13~{\rm TeV}$ LHC
\begin{equation}
\sigma^{\rm LO}_{\rm VBF} = 957~{\rm fb},\;\;\;\;
\sigma^{\rm NNLO, NF}_{\rm VBF} = -3.73~{\rm fb}\,,
\end{equation}
which implies a negative  nonfactorizable correction
\begin{equation}
  \Delta_{\rm NF} =
  \frac{\sigma^{\rm NNLO, NF}_{\rm VBF}}{\sigma^{\rm LO}_{\rm VBF}}
  \times 100 \% = -0.39\% \,.
  \label{eq:corr}
\end{equation}
While the nonfactorizable   correction is small, it is quite comparable to
the  N$^3$LO QCD factorizable corrections computed in
Ref.~\cite{Dreyer:2016oyx}.  We note that the choice of a proper
renormalization scale for the computation of nonfactorizable corrections is
an interesting problem. Indeed, as follows from our computation, they appear
for the first time at NNLO and so their scale dependence is not  compensated.
If we simply  decrease (increase) the renormalization scale in
Eq.~(\ref{eq:scales}) by a factor of two, $\Delta_{\rm NF}$ changes to
$-0.5\%$ ($-0.3\%$), respectively.

The situation becomes even more interesting   when we consider differential
distributions.  For example, in Fig.~\ref{fig::3} the nonfactorizable  QCD
corrections to the transverse momentum distributions of the two jets and the
Higgs boson as well as various rapidity distributions are shown. For each
plot, the upper panel displays leading order distributions whereas the lower
panel shows the
correction $\Delta_{\rm NF}$, {\it cf.}
Eq.~(\ref{eq:corr}), in dependence of a relevant kinematic variable.  As it follows from the plots, the corrections to the jet
transverse momenta distributions depend strongly on $p_{j,\perp}$  and can
even exceed $1\%$ in certain cases.
By contrast, the correction to Higgs
transverse momentum is rather flat and, for this reason, is comparable to the
correction to the VBF cross section Eq.~(\ref{eq:corr}).  Correction to
the rapidity distribution of the Higgs boson is rather flat too but some
dependence on the rapidity is present in the corrections to the leading jet
rapidity distribution and to the distribution in the rapidity difference of
the two jets. The correction to the rapidity distribution of the second jet
is similar to that of the first and we therefore do not show it separately.

We emphasize that in the numerical simulation we keep the full dependence of the
leading order cross section on the kinematic variables without expanding in
transverse momenta.  The plots in  Figs.~\ref{fig::3} show that kinematic
distributions peak at  $p_{j,\perp}\approx 100$~GeV, which suggests
that the leading ${\cal O} (p^2_{j,\perp}/s)$ power corrections  to our
result for $\sigma^{\rm NNLO}_{\rm nf}$ are in a few percent range and, for
this reason, negligible.

Thus we have obtained  analytic results for the nonfactorizable NNLO QCD
corrections  to the Higgs boson production in the vector boson fusion valid
in the phenomenologically  most interesting kinematic region where the
characteristic transverse momenta  are  much smaller than the center-of-mass
energy of the process and a rapidity gap between the Higgs boson and
the tagging jets is present.  The leading in $p_{j,\perp}/\sqrt{s}$ correction is
related to the Glauber phase and has a natural $\pi^2$-enhancement along with
the  color suppression relative to the factorizable ones. It exhibits
nontrivial dependence on the transverse momenta and rapidities of the
tagging jets. Numerically, the corrections are found to be close to  half of a percent
although they can become as large as a percent  in certain kinematic regions.

{\bf Acknowledgments.} We are grateful to M.~Schulze and F.~Caola for
providing us with the numerical code for  Higgs boson production in vector
boson fusion. The research of T.L. was supported by NSERC. The research of
K.M. and A.P.  was supported by the Deutsche Forschungsgemeinschaft (DFG,
German Research Foundation) under grant 396021762 - TRR 257.  The research of
A.P. was supported by NSERC and Perimeter Institute for Theoretical
Physics.

%%%%%%%%%%%%%%%%%%%%%%%%%%%%%%%%%%%%%%%%%%%%%%%%%%%%%%%%%%%%

\end{document}